\documentclass{article}
\usepackage{spconf,amsmath,graphicx}
\usepackage{hyperref,amssymb,ctable,makecell,multirow,multicol,subcaption}

\title{Learned Transferable Architectures Can Surpass Hand-Designed Architectures for Large Scale Speech Recognition}
\name{Liqiang He$^1$, Dan Su$^1$, Dong Yu$^2$}
\address{
  $^1$Tencent AI Lab, Shenzhen, China\\
  $^2$Tencent AI Lab, Bellevue WA, USA}
\begin{document}
\ninept
\maketitle
\begin{abstract}
In this paper, we explore the neural architecture search (NAS) for automatic speech recognition (ASR) systems. We conduct the architecture search on the small proxy dataset, and then evaluate the network, constructed from the searched architecture, on the large dataset. Specially, we propose a revised search space that theoretically facilitates the search algorithm to explore the architectures with low complexity. Extensive experiments show that: (i) the architecture learned in the revised search space can greatly reduce the computational overhead and GPU memory usage with mild performance degradation. (ii) the searched architecture can achieve more than 15\% (average on the four test sets) relative improvements on the large dataset, compared with our best hand-designed DFSMN-SAN architecture. To the best of our knowledge, this is the first report of NAS results with a large scale dataset (up to 10K hours), indicating the promising application of NAS to industrial ASR systems.
\end{abstract}
\section{Introduction}
\label{sec:intro}

The performance of ASR systems has been largely boosted by deep learning \cite{hinton2012deep}. The core part of deep learning is to design and optimize deep neural networks. Various types of neural network architectures have been employed in ASR systems, such as convolutional neural networks (CNNs) \cite{sainath2013deep}, long short-term memory (LSTM) \cite{graves2013hybrid}, gated recurrent unit\cite{ravanelli2018light}, time-delayed neural network \cite{peddinti2015time},  feedforward sequential memory networks (FSMN) \cite{zhang2018deep}, etc.
Some combinations of different architectures are also proposed to take advantage of their complementary property, such as CLDNN \cite{sainath2015convolutional}. Recently, transformer architecture which has achieved its success in the natural language process (NLP) tasks has also been widely used in ASR systems \cite{dong2018speech, pham2019very}, demonstrating its superior performance compared with the state-of-the-art models. Our previous work also proposed a variant of model architecture which combined DFSMN with self-attention networks (SAN), and further applied the memory augmenting method on the self-attention layer \cite{you2020dfsmn}. In summary, the performance improvement of ASR systems owes much to the dedicated hand-designed model architectures.

However, designing state-of-the-art neural network architectures requires a lot of expert knowledge and takes ample time. Therefore, there has been a growing interest in developing algorithmic solutions to discover powerful network architectures automatically. The network architectures automatically searched in \cite{zoph2016neural, zoph2018learning, real2019regularized} have achieved highly competitive performance in computer vision tasks, such as image classification and object detection. But, the heuristic search methods with evolution and reinforcement learning technique require massive computational overheads (3150 GPU days of evolution \cite{real2019regularized} and 2000 GPU days of reinforcement learning \cite{zoph2018learning}). Several approaches focusing on the efficient architecture search have been proposed.
Among them, DARTS \cite{liu2018darts} introduced a differentiable NAS framework to relax the discrete search space into a continuous one by weighting candidate operations with architectural parameters, which achieved comparable performance and remarkable efficiency improvement compared to previous approaches. As a progressive version of DARTS, P-DARTS \cite{chen2019progressive} was proposed to bridge the depth gap between the network depth of architecture search and architecture evaluation.

\begin{figure}[tp!]
    \centering
    \includegraphics[width=0.70\linewidth]{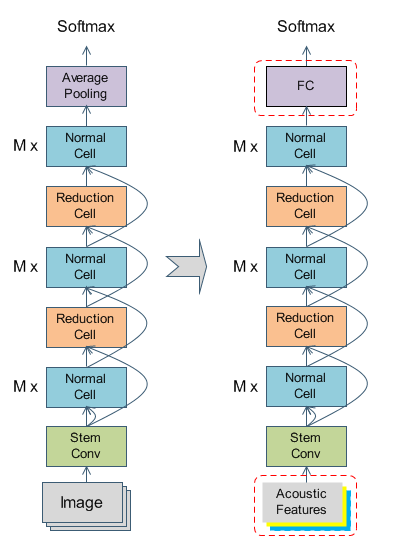}
    \vspace{-0.1cm}
    \caption{The convolutional architecture (left) for computer vision task, and the convolutional architecture (right) for speech recognition task, as proposed by our work. Abbreviation: M refers to the number of normal cells, FC refers to the fully connected layers.}
    \label{fig:arch}
    \vspace{-0.3cm}
\end{figure}

Despite the rapid advance of NAS techniques in the computer vision communities, there has been very limited research on the application of NAS to ASR systems. Compared with computer vision tasks such as image classification, a major hindrance is that speech recognition is a more complex task in terms of the dimension of the input and output. What's more, for the big data era of speech recognition, a typical amount of training data can be over 10K hours, which amounts to more than 10 million samples.

In this work, we explore the feasibility of NAS for speech recognition tasks on the large dataset. Considering the search cost, we make the architecture search with multiple stages following P-DARTS on the small proxy dataset, and then the evaluation network, constructed from the searched architecture, is evaluated on the large dataset. We propose a revised search space for speech recognition tasks which theoretically facilitates the search algorithm to explore the architectures with low complexity, compared with the DARTS-based search space. Experimental results show that the architecture, discovered in the revised search space on the AISHELL-1 dataset, can achieve more than 20\% and 15\% (average on the four test sets) relative improvements respectively on the AISHELL-2 dataset and the large (10k hours) dataset, compared with our best hand-designed DFSMN-SAN architecture.

Our contributions can be summarized as follows:
(i) We show that the architectures searched on the small proxy dataset has good transferability to the large (10k hours) dataset for the speech recognition tasks.
(ii) We propose a revised search space, from which the searched architecture achieves a better balance between model complexity and recognition performance.
(iii) We show that the searched architecture achieves significant performance improvements on the large dataset, compared with our best hand-designed model architecture.


\section{Neural Architecture Search}
\label{sec:2}

\subsection{DARTS \cite{liu2018darts}}

Different from conventional methods applying evolution or reinforcement learning over a discrete search space, a differentiable network architecture search based on bilevel optimization is introduced in DARTS, which achieves remarkable efficiency improvement by several orders of magnitude. The categorical choice of one operation is relaxed to learning a set of continuous variables $\alpha=\{\alpha^{(i,j)}\}$, normalized with the \textsl{Softmax} function.
\begin{align}
\label{eq:8}
\overline{o}^{(i,j)}(x)=\sum\limits_{o\in\mathcal{O}}\frac{exp(\alpha_o^{(i,j)})}{\sum_{o^{\prime}\in\mathcal{O}}exp(\alpha_{o^{\prime}}^{(i,j)})}o(x)
\end{align}

A bilevel optimization is proposed which jointly optimizes the architecture $\alpha$ as the upper-level variable and the network weights $\mathcal{\omega}$ as the lower-level variable:
\begin{align}
\label{eq:8}
&\mathop{min}\limits_{\alpha} \mathcal{L}_{val}(\mathcal{\omega}^{*}(\alpha), \alpha) \\
&s.t.\quad\mathcal{\omega}^{*}(\alpha)=\mathop{argmin}_{\mathcal{\omega}}\mathcal{L}_{train}(\omega, \alpha) 
\end{align}
where $\mathcal{L}_{train}$ and $\mathcal{L}_{val}$ denote the training and the validation loss, respectively. Both losses are determined not only by the architecture $\alpha$, but also the network weights $\mathcal{\omega}$.

\subsection{P-DARTS \cite{chen2019progressive}}

Although good transferability of the searched architecture has been observed in DARTS, more attention has been paid to the discrepancies between the \textsl{super-network} (the continuous architecture encoding) and the evaluation network constructed from the optimal \textsl{sub-network} (the derived discrete architecture) for the specific task. One of the discrepancies is the \textsl{depth gap} between the depth of the super-network and the evaluation network, which has been proven to cause performance deterioration. As a progressive version of DARTS, \textsl{search space approximation} is proposed in P-DARTS to alleviate the problem of the depth gap by dividing the search process into multiple stages. With each stage forward, the depth of the super-network becomes deeper, while at the same time the number of operations in the search space becomes smaller, which makes the search process with a deeper super-network possible with the limited computation and memory budget. Additionally, \textsl{search space regularization} is introduced to address the ``over-fitting'' problem brought by the \textsl{skip-connect} operation.

\begin{figure*}[tp!]
    \centering
    \vspace{-0.1cm}
    \begin{subfigure}{0.48\textwidth}
         \centering
         \includegraphics[width=0.96\textwidth]{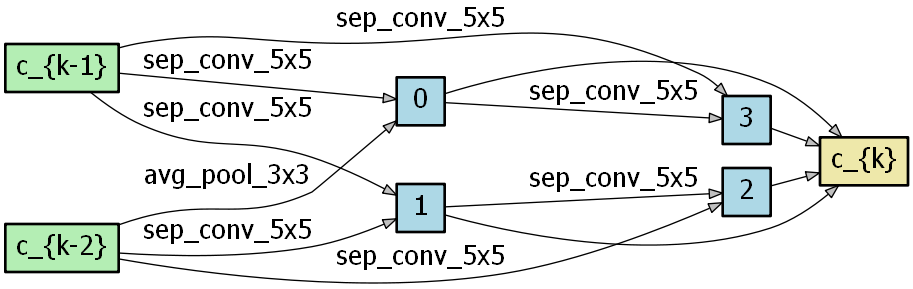}
         \caption{Normal cell learned in original search space.}
         \label{subfigure-1}
    \end{subfigure}
    \hfill
    \begin{subfigure}{0.48\textwidth}
         \centering
         \includegraphics[width=0.96\textwidth]{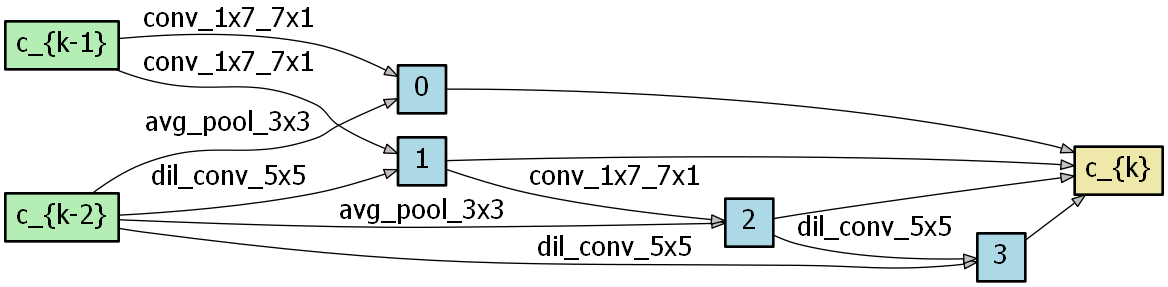}
         \caption{Normal cell learned in revised search space.}
         \label{subfigure-2}
    \end{subfigure}
    \vfill
    \begin{subfigure}{0.48\textwidth}
         \centering
         \includegraphics[width=0.96\textwidth]{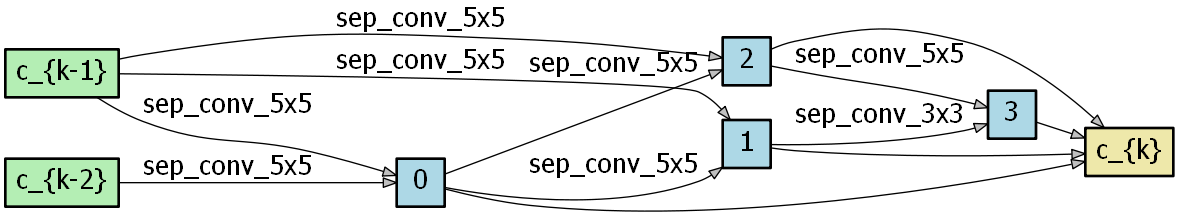}
         \caption{Reduction cell learned in original search space.}
         \label{subfigure-3}
    \end{subfigure}
    \hfill
    \begin{subfigure}{0.48\textwidth}
         \centering
         \includegraphics[width=0.96\textwidth]{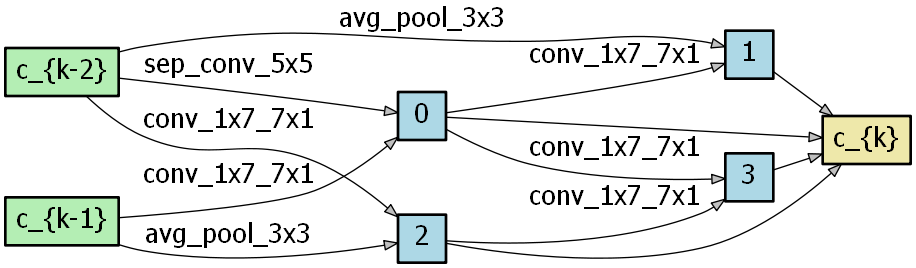}
         \caption{Reduction cell learned in revised search space.}
         \label{subfigure-4}
    \end{subfigure}
\vspace{-0.1cm}
\caption{(a) and (c) are the cells (denoted as ASRNET-A) learned in the original search space. (b) and (d) are the cells (denoted as ASRNET-B) learned in the revised search space proposed by our work.}
\label{fig:cell}
\vspace{-0.3cm}
\end{figure*}

\section{Search Space Revision}
\label{sec:3}

In this section, we summarize the characteristics and the improvements of the search algorithm when applying NAS to ASR systems. Based on the architecture in DARTS, there are two modifications specifically made for the speech recognition tasks. For large scale speech recognition, we propose a revised search space that theoretically facilitates the search algorithm to explore the architectures with low computational and memory overhead. Besides, the regularization method for the architecture search is discussed.

\subsection{Architecture for speech recognition}
\label{sec:arch}

Following DARTS, we search for the convolutional cells as the building blocks and then stack the learned cells to form the final network architecture. As shown in Figure \ref{fig:arch} (left), each cell connects to the previous two cells or stem convolutions located at the beginning of the network. Cells located at the 1/3 and 2/3 of the network are reduction cells (totally two), which is different from the other normal cells.
Two modifications for speech recognition are made as shown in the red dotted boxes of Figure \ref{fig:arch} (right). In the modification at the beginning of the network, acoustic features and the first-order and the second-order derivatives are separately assigned to the independent channels. In the modification at the ending of the network, the \textsl{average pooling} operation is replaced by several fully connected layers, following with the \textsl{softmax} layer to compute the posteriors. Moreover, there are two reduction cells in the network, each of which reduces the resolution of the feature maps from the previous cells by half, and this architecture is also adopted for the speech recognition task, as the lower frame rate technique proposed by \cite{golan2016rate} has shown its benefit.

\subsection{Search Space Revision and Regularization}
\label{sec:revised}

The search space for the speech recognition task is represented in the form of the convolutional cells, and each cell is denoted as a directed acyclic graph consisting of N nodes and their corresponding edges. Each directed edge between two nodes is associated with the candidate operations. In DARTS, The candidate set in the convolutional cells includes the following operations: \textbf{\emph{[zero, identity, 3x3 max pooling, 3x3 average pooling, 3x3 separable convolution, 5x5 separable convolution, 3x3 dilated separable convolution,  5x5 dilated separable convolution]}}.

Considering the depth gap between the network depth applied in the search and the evaluation for the speech recognition tasks, the search process of our work adopts the search space approximation method proposed by P-DARTS. Based on the DARTS-based operation space, the preliminary architecture searches are carried out on the proxy dataset. One of the searched architecture is shown in Figure \ref{fig:cell} (left).

We explore three issues related to the search space when applying P-DARTS for speech recognition tasks.
\textbf{First}, the search process of the computer vision tasks tends to generate architectures with many \textsl{skip-connect} operations, especially on the small proxy dataset, and the derived architectures for evaluation often suffer from the performance degradation.
But, after searching on a small proxy dataset, we find that the \textsl{skip-connect} operation rarely appears in the final searched architecture for the speech recognition tasks. This is arguably due to that speech recognition is a more complex task in terms of both the dimension of the input and output, compared with computer vision tasks such as image classification.
\textbf{Second}, the cell architectures, learned in the DARTS-based search space, with better performance are prone to have many \textsl{separable convolution} operations, and this operation applies the module list twice, the module list that consists of sequential modules with a \textsl{ReLU-Conv-BN} order. However, such learned architectures applied for large scale speech recognition consumes too much computation resources and GPU memory, which can be prohibitive due to the limitation of GPU hardware. 
\textbf{Last}, to eliminate the influence of randomness, the search process in DARTS and P-DARTS should be repeated several times with different seeds for the final architecture with better performance, and this method is still applicable for the search process of the speech recognition. Notably, the search process of the speech recognition tends to generate architectures with many \textsl{average pooling} and \textsl{dilated convolution} operations, and obvious performance fluctuations have been observed based on the evaluation networks derived by stacking the learned normal cells for more times.

Concerning the first two issues, we revise the operation space by replacing \textsl{skip-connect} operation with \textsl{1x7 then 7x1} convolution \cite{real2019regularized}. The \textsl{skip-connect} operation has lower priority in the relaxed search space for the absence in the final architectures most of the time, so the stability of the search process can be improved by removing this operation. The newly added convolution operation has the following advantages. First, the convolution with the larger convolution kernel increases the receptive field to capture the latent representation of acoustic features and meanwhile limits the number of model parameters as smaller as possible. Second, the convolution with fewer sequential modules improves computing efficiency and memory overhead. With the revised operation space, the architecture searches for the speech recognition tasks are carried out on the small proxy dataset, and one of the searched architecture is shown in Figure \ref{fig:cell} (right). The evaluation network, constructed from the searched architecture, achieves a better trade-off between model complexity and recognition performance. The revised operation space in the convolutional cell includes the following operations: \textbf{\emph{[zero, 3x3 max pooling, 3x3 average pooling, 3x3 separable convolution, 5x5 separable convolution, 3x3 dilated separable convolution,  5x5 dilated separable convolution, 1x7 then 7x1 convolution]}}.

As for the last issue mentioned above, we adopt the search space regularization proposed by \cite{chen2019progressive} to alleviate the problem of obvious performance fluctuations caused by the randomness of the search process. First, the operation-level \textsl{dropout} is inserted after each \textsl{dilated separable convolution} and \textsl{average pooling} operation to facilitate the algorithm to explore other operations. Second, the regularization rule of architecture refinement restricts the number of preserved \textsl{average pooling} operations in the final architecture to be a constant. \textsl{average pooling} operation has a significant impact on recognition accuracy at the evaluation stage, especially on the proxy dataset with complex scenarios.

\section{Experiments}
\label{sec:4}

\subsection{Datasets}

We use AISHELL-1 \cite{bu2017aishell} as the small proxy dataset for the architecture search. The dataset contains 178 hours of Chinese Mandarin speech from 400 speakers, and the 10 hours test set is used for the architecture evaluation. Two bigger corpora are used to verify the transferability of the searched architecture. First is the AISHELL-2 \cite{du2018aishell} dataset which contains 1000 hours of speech data from 1991 speakers. Second is a 10K hours multi-domain dataset \cite{you2020dfsmn}. We also augment the  AISEHLL-1 and AISEHLL-2 training data with 2-fold speed perturbation \cite{ko2015audio} in the experiments. To evaluate the performance of the searched architecture, we report performance on 3 types of test sets which consist of hand-transcribed anonymized utterances extracted from reading speech (1001 utterances), conversation speech (1538 utterances), and spontaneous speech (2952 utterances). We refer them as Read, Chat, and Spon respectively. Besides, to provide a public benchmark, we also use the AISHELL-2 development set (2500 utterances, short for DEV) recorded by high fidelity microphone as the test set.

\subsection{Training setup}

We use 40-dimensional log Mel-filterbank features with the first-order and the second-order derivatives. Training utterances are filtered by a maximum frame length of 1024, and the length of each utterance is padded to be 4-frames-aligned to be fit for the two reduction layers. All the experiments are based on the CTC learning framework and trained with multiple GPUs using BMUF \cite{chen2016scalable} optimization. We use CI-syllable-based acoustic modeling units which include 1394 Mandarin syllables, 39 English phones, and a blank. First-pass decoding with a pruned 5-gram language model is performed with a beam search algorithm by using the weighted finite-state transducers (WFSTs). Character error rate (CER) results are measured on the test sets. \textsl{Rel Imp} refers to \textsl{Relative Improvement} in Table \ref{tab:medium} and Table \ref{tab:large}.

\begin{table}[tp!]
\centering
\caption{Token accuracies (Acc) and evaluation costs (Cost) of the small evaluation networks on the AISHELL-1 dataset. Abbreviations: L is the number of cells, C is the initial number of channels.}
\label{tab:small}
\vspace{-0.1cm}
\begin{tabular}{c|c|c|c}
\specialrule{.13em}{0em}{0em}
\makecell{Small} & \makecell{Params\\($M$)} & \makecell{Acc\\($\%$)} & \makecell{Cost\\($hours$)} \\
\hline
\makecell{ASRNET-A (L=17,C=32)} & \makecell{6.6} & \makecell{93.00} & \makecell{35.2} \\
\hline
\makecell{ASRNET-B (L=17,C=24)} & \makecell{6.8} & \makecell{92.24} & \makecell{17.8} \\
\specialrule{.13em}{0em}{0em} 
\end{tabular}
\vspace{0.1cm}
\end{table}

\begin{table}[tp!]
\centering
\caption{Comparison with DFSMN-SAN architecture on the AISHELL-2 dataset.}
\label{tab:medium}
\vspace{-0.1cm}
\begin{tabular}{c|c|c|c}
\specialrule{.13em}{0em}{0em}
\makecell{Medium} & \makecell{Params\\($M$)} & \makecell{CER\\($\%$)} & \makecell{\textsl{Rel Imp} \\($\%$)}\\
\hline
\makecell{DFSMN-SAN} & \makecell{14.4} & \makecell{7.36} &  - \\
\hline
\makecell{ASRNET-A (L=32,C=38)} & \makecell{12.1} & \makecell{5.65} & \makecell{\textsl{23.2}}  \\
\hline
\makecell{ASRNET-B (L=32,C=30)} & \makecell{14.7} & \makecell{5.79} & \makecell{\textsl{21.3}}\\
\specialrule{.13em}{0em}{0em} 
\end{tabular}
\vspace{-0.3cm}
\end{table}

\subsection{Architecture Search}

The training set of the AISHELL-1 dataset is randomly split into two equal subsets, one for learning network parameters and the other for tuning the architectural parameters. The search process following \cite{chen2019progressive} is divided into three stages. For each stage, the super-network is trained for 15 epochs, with batch size 4 (for both the training and validation sets) and the initial number of channels 16. Only network parameters are trained in the first 6 epochs, and both network and architecture parameters are alternately optimized in the rest 9 epochs. The momentum SGD optimizer with initial learning rate 0.01 (annealed down to zero following a cosine schedule without restart), momentum 0.9, weight decay 0.0003, is adopted to optimize the network parameters. The dropout probability on \textsl{dilated separable convolution} and \textsl{average pooling} is decayed exponentially and the initial values are set to be 0.05, 0.05, 0.05 for stage 1, 2 and 3, respectively. The final discovered normal cells are restricted to keep at most 2 \textsl{average pooling}. We run the search processes separately in the original (DARTS-based) search space and the revised search space proposed by our work. Concerning the influence of randomness, the search process is repeated 3 times with different seeds, respectively for both the DARTS-based search space and the revised search space. The search process takes around 89 hours on 8 Tesla P40 GPUs.

\subsection{Architecture Evaluation}

On the AISHELL-1 dataset, the small evaluation networks stacked \cite{zoph2018learning} with 17 cells are trained from scratch for 20 epochs with batch size 4. Other hyper-parameters remain the same as the ones used for the architecture search. Based on the recognition performance on the test set of AISHELL-1 dataset, the final architecture (denoted as \textsl{ASRNET-A}) discovered in the original search space is shown in Figure \ref{fig:cell} (left) and the final one (denoted as \textsl{ASRNET-B}) discovered in the revised search space is shown in Figure \ref{fig:cell} (right). The initial numbers of channels are 32, 24 for \textsl{ASRNET-A} and \textsl{ASRNET-B} respectively. The token accuracies are computed on the test set of the AISHELL-1 dataset. As seen in Table \ref{tab:small}, the performance of the evaluation network constructed from \textsl{ASRNET-A} is slightly better than the one constructed from \textsl{ASRNET-B}, but the training time of the former almost takes almost twice as long as the latter. The training tasks are performed on 8 Tesla P40 GPUs.

To test the transferability of the searched architectures, the medium-sized evaluation networks stacked with 32 cells are trained from scratch for 15 epochs with batch size 4, on the AISHELL-2 dataset. The initial numbers of channels are 38, 30 for \textsl{ASRNET-A} and \textsl{ASRNET-B} respectively. Other training configurations are the same as the ones used for the small evaluation network. Character error rate results are computed on the test set of the AISHELL-2 dataset. As shown in Table \ref{tab:medium}, the medium-sized networks have achieved more than 20\% relative improvements, compared with \textsl{DFSMN-SAN} \cite{you2020dfsmn} consisting of 10 \textsl{DFSMN} components and 2 multi-head self-attention sub-layers. Concerning computational overhead and GPU memory usage, the evaluation network constructed from \textsl{ASRNET-A} is almost twice as much as the one constructed from \textsl{ASRNET-B}, so the latter achieves a better trade-off between model complexity and recognition performance. The training processes are accelerated by applying 24 Tesla P40 GPUs.

To further validate the transferability of the searched architecture \textsl{ASRNET-B}, the large evaluation network stacked with 32 cells is trained from scratch for 6 epochs with batch size 4 and the initial number of channels 50, on the 10K hours large dataset. The momentum SGD optimizer with initial learning rate 0.0002, momentum 0.9, weight decay 0.0003, is adopted to optimize the network parameters. As shown in Table \ref{tab:large}, compared with \textsl{DFSMN-SAN} consisting of 30 \textsl{DFSMN} components and 3 multi-head self-attention sub-layers, the large network has achieved more than 15\% (average on the four test sets) relative improvements. The significant performance improvements imply that the searched architecture has good transferability. The training process takes around 7 days on 24 Tesla V100 GPUs.

\begin{table}[tp!]
\centering
\caption{Comparison with DFSMN-SAN architecture on the 10k hours dataset. CERs are measured on the four test sets.
}
\label{tab:large}
\vspace{-0.1cm}
\begin{tabular}{c|c|c|c|c|c}
\specialrule{.13em}{0em}{0em}
\makecell{Large} & \makecell{Params\\($M$)} & \makecell{Read\\(\%)} & \makecell{Chat\\(\%)} & \makecell{Spon\\(\%)} & \makecell{DEV\\(\%)} \\
\hline
\makecell{DFSMN-SAN} & \makecell{36.1} & \makecell{1.95} & \makecell{22.92} & \makecell{25.41} & \makecell{4.42} \\
\hline
\makecell{ASRNET-B\\(L=32,C=50)} & \makecell{36.7} & \makecell{1.61} & \makecell{19.99} & \makecell{20.83} & \makecell{3.86} \\
\hline
\makecell{\textsl{Rel Imp}} & \makecell{-} & \makecell{\textsl{17.38}} & \makecell{\textsl{12.79}} & \makecell{\textsl{18.05}} & \makecell{\textsl{12.58}} \\
\specialrule{.13em}{0em}{0em} 
\end{tabular}
\vspace{-0.3cm}
\end{table}

\section{Conclusions}
\label{sec:5}

In this paper, we empirically show that not only is the application of NAS for large scale acoustic modeling in speech recognition possible, but it also allows for very strong performance. Specifically, we perform the architecture search on 150 hours small dataset and then transfer the searched architecture to a large dataset for evaluation. On the 1000 hours AIShell-2 and 10K hours multi-domain datasets, the searched architecture achieves more than 20\% and 15\% (average on the four test sets) relative improvements respectively compared with our best hand-designed model architecture. The study of this work may unleash the potentials of NAS application for ASR systems. Future work includes adding latency control constraints into NAS to perform the architecture search for streaming ASR scenarios.

\bibliographystyle{IEEEbib}
\bibliography{strings,refs}

\end{document}